# GAS TRAP PREDICTION FROM 3D SEISMIC AND WELL TEST DATA USING MACHINE LEARNING

**Dmitry Ivlev**

dm.ivlev@gmail.com

**The aim** of this work is to create and apply a methodological approach for predicting gas traps from 3D seismic data and gas well testing.

**Materials and methods.** Data from 12 drilled wells, 1368 seismic field attributes, covering an area of 456 km$^2$ in the TVD depth interval from 1200 to 2400 m were used in this work. The proposed approach is based on binary classification algorithms with training on well data. The study includes the following sequence of operations: interpretation of gas well test results to determine the radius of radial gas filtration (IARF); correlation of the top and bottom of productive horizons based on seismic data in the near-wellbore space; creating volumes of space for positive and negative class; dividing the sample into training and validation; creation of a separate test sample for the metamodel; creation of a feature space - extraction of seismic wavefield attributes; creation of data sampling - assignment of a vector of seismic attributes to each point of space within the volumes of classes; selection of features; basic model training; generalized assessment of trait contribution; creation of an ensemble of classification models using a metamodel - logistic regression; prediction of the probability of space belonging to gas reservoirs; evaluation of forecast quality on a test sample.

**Conclusion.** The paper formalizes an approach to creating a training dataset by selecting volumes with established gas saturation and filtration properties within a seismic wavefield. The volumes divide the studied space into positive and negative classes. Positive class is a volume of gas-saturated sands, identified by the results of detailed correlation of the gas sandstone top and bottom, within which there is a region with a boundary along the radius of continuous radial gas filtration. The radius was estimated by interpreting gas well test data after deconvolution of test regimes. The negative class includes volumes of explored space within a one-hundred-meter radius along the wellbore with proven futility based on test results and log interpretation. Each point in the space within the classes is assigned seismic field attributes. The training data set created in this way is used in the technological stack of sequential application of data processing methods and ensemble machine learning algorithms.

**As a result,** the cube of calibrated probabilities of belonging of the studied space to gas reservoirs is obtained. The high efficiency of this approach is shown on a delayed test sample from three wells (blind wells). The final value of the gas sandstone prediction quality metric f1 score was 0.893846.

**Keywords:** machine learning, well data, seismic attributes, facies prediction, rock properties prediction, augmentation methods, ensemble learning, feature selection, evaluation of feature contribution to prediction, geophysics, well testing, deconvolution of well testing.

**Introduction**

Well testing and 3D seismic surveys are standard surveys in the oil and gas industry. Information from these types of surveys provides insight into the characteristics of the rock in the interwell space. The surveys utilize different physical principles to obtain information. Measurements of pressure in the perforation interval and fluid withdrawal volumes during well testing can reconstruct fluid filtration conditions and properties over a radius of distance from the pressure transducer. Seismic exploration uses seismoacoustic waves over the entire fieldwork area to study geologic formations at depth. Seismic surveys can cover significant areas of space, from units to thousands of square kilometers. The area of investigation during well testing is determined by the spread of the pressure front in the pay zone within the radius of the well being tested. Depending on the time of investigation and the influence of boundary effects on filtration, this area can reach up to one square kilometer over the entire thickness of the productive interval, which makes well testing a method of investigating productive horizons comparable in scale to seismic studies. When these methods are conducted sequentially in the same area, their study areas overlap. Thus, with a certain degree of reliability, it is possible to identify volumes in seismic data with established HC saturation and filtration properties. Information on productive and unproductive volumes can be used in the task of binary classification with a teacher, when a continuous function of the probability of hydrocarbon saturation of the rock over the entire 3D seismic survey area is reconstructed using machine learning algorithms on the attributes of the seismic field.

Accordingly, the aim of this paper is to predict gas traps from 3D seismic and gas well test data using a technology stack of data processing algorithms and machine learning ensembles. The study includes the following sequence of operations: interpretation of gas well test results to determine the radius of radial gas filtration (IARF); correlation of the top and bottom of productive horizons based on seismic data in the near-wellbore space; creating volumes of space for positive and negative class; dividing the sample into training and validation; creation of a separate test sample for the metamodel; creation of a feature space - extraction of seismic wavefield attributes; creation of data sampling - assignment of a vector of seismic attributes to each point of space within the volumes of classes; selection of features; basic model training; generalized assessment of trait contribution; creation of an ensemble of classification models using a metamodel - logistic regression; prediction of the probability of space belonging to gas reservoirs; evaluation of forecast quality on a test sample.

**Object of study**

The study area is located in a land area in North Africa covered by 3D seismic surveys and wells. The object of the study is productive gas-bearing sediments, represented by sands, near the base of the Kafr El Sheikh (KES) Formation in the interval of its spreading across the study area.

**Input data**

The work area is covered by 3D seismic surveys with an area of 456 km$^2$. According to the results of applying different processing graph, 4 cubes of seismic amplitudes in depths with horizontal resolution of 25 meters and vertical resolution of 5 meters were obtained. To form the survey space, an area equal to the seismic survey area in the TVD depth interval from 1200 to 2400 meters was cut out of the cubes.

Twelve vertical wells were drilled in the seismic exploration area, nine of which produced commercial gas flows with minor amounts of condensate. At the time of analysis, cased hole gas horizon testing was performed in 7 wells after drilling. Logging data in all wells determined the presence or absence of productive targets in the drilled interval.

**Interpreting gas well testing**

Interpretation of gas well testing was performed in two stages. In the first step, deconvolution of the test regimes was performed. Deconvolution converts the variable-flow pressure data into an initial constant-flow decline with a duration equal to the total test duration and directly yields the corresponding pressure derivative normalized to the unit flow rate. This derivative is thus free from distortions caused by the pressure derivative calculation algorithm and from errors caused by incomplete or truncated flow rate histories [1].

At the second stage, using the deconvolution curve on the Bourdet derivative, the radial filtration regime (IARF) was identified, the filtration-capacitance properties of the productive interval were calculated, and the end time of the radial gas filtration regime was estimated.

It does not matter what boundary effect met the pressure change front, only the volume of rock through which continuous radial gas filtration was carried out is important for this work.

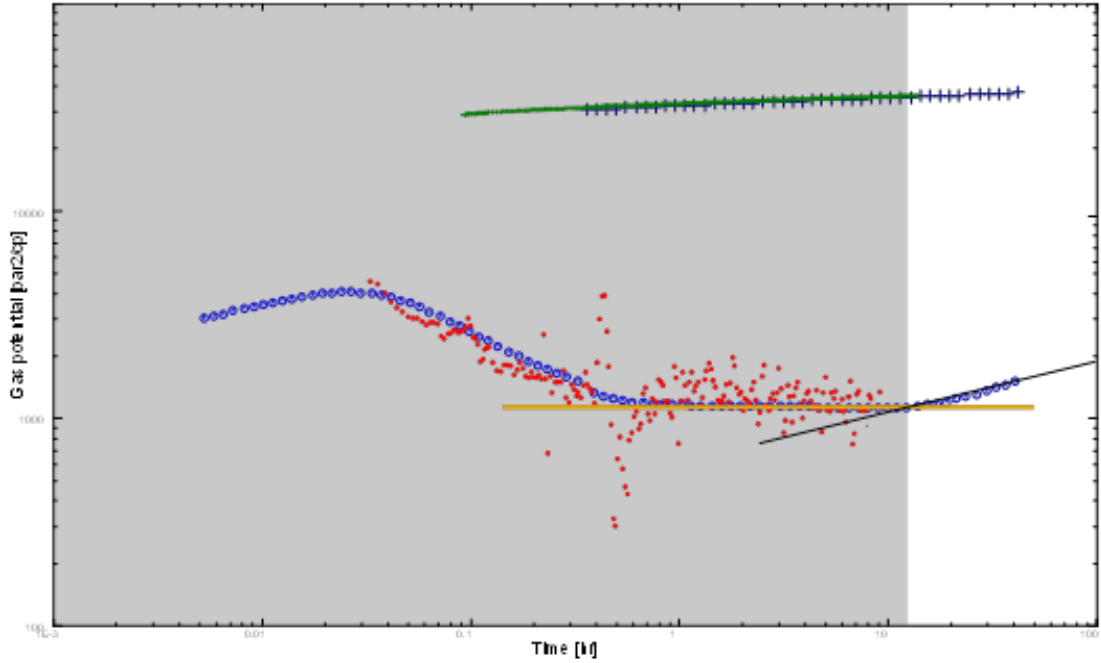

Figure 1: Graph in double logarithmic scale. Interpretation of the IY-1X gas well test. Red dots - Bourdais derivative of the longest pressure recovery period. Green dots - pressure dynamics of this regime. Blue dots - derivative of the deconvolution curve. Blue crosses - dynamics of the deconvolution curve. Yellow line - approximation on the data of the derivative, straight line with zero slope. Black straight line - approximation of the straight line on the deconvolution data, related to the boundary effect of plane-radial flow. Gray area - interval of flat-radial flow in the productive interval accepted for further study.

The initial radial flow filtration distance is determined based on the following formula:

$$r_{IARF} = 0.029 \sqrt{\frac{k \, t_{IARF}}{\varphi \, \mu \, c_t}} \quad ,$$

where $k$ is the rock permeability, $t_{IARF}$ is the time of the end of the flat-radial fluid filtration mode, $\varphi$ porosity, $\mu$ fluid viscosity, $c_t$ total compressibility of the system [2].

According to the results of interpretation the IARF radii of gas in the productive horizon were obtained (Table 1).

Table 1. Radii of radial filtration of gas in the productive horizon

| Well | $r_{IARF}$, м |
|---|---|
| IY-1X | 106 |
| IY-2 | 221 |
| MA-1X | 114 |
| SD-12X | 62 |
| SD-1X | 445 |
| SD-3X | 154 |
| SD-4X | 434 |

**Creating classes**

Detailed correlation of seismic horizons in the near-wellbore space was performed to identify class volumes in the seismic data. The correlated surfaces were approximated to the upper and lower boundaries of the productive interval. Thus, volumes bounded by surfaces and radius of the radial filtration mode were obtained. The created volumes of gas sandstone formed spaces in the seismic data for the positive class, which is shown on the section by the area with yellow shading (Figure 2a).

Two approaches were used to generate volumes of negative class label space.

For the wells in which gas tests were conducted, two volumes were created in the form of cylinders with a radius of 100 meters centered on the wellbore: the first volume was bounded from below by an amplitude-correlated surface with an indentation from the boundary of the productive interval of 30 meters and from above by the upper boundary of the study; the second volume was bounded from above by a surface with the same indentation and from below by an indentation from the boundary of the productive interval of 30 meters; the upper one was bounded from above by the depth of the study, the lower one by the depth of the well; the radius of the cylinder was bounded from above by the depth of the well (Figure 2a).

For wells that did not penetrate the productive horizon, a cylinder was created along the wellbore with a radius of 100 meters, limited by the depth of the well from below and the upper boundary of the study (Figure 2b). Thus, 10 wells were divided into productive and non-productive classes.

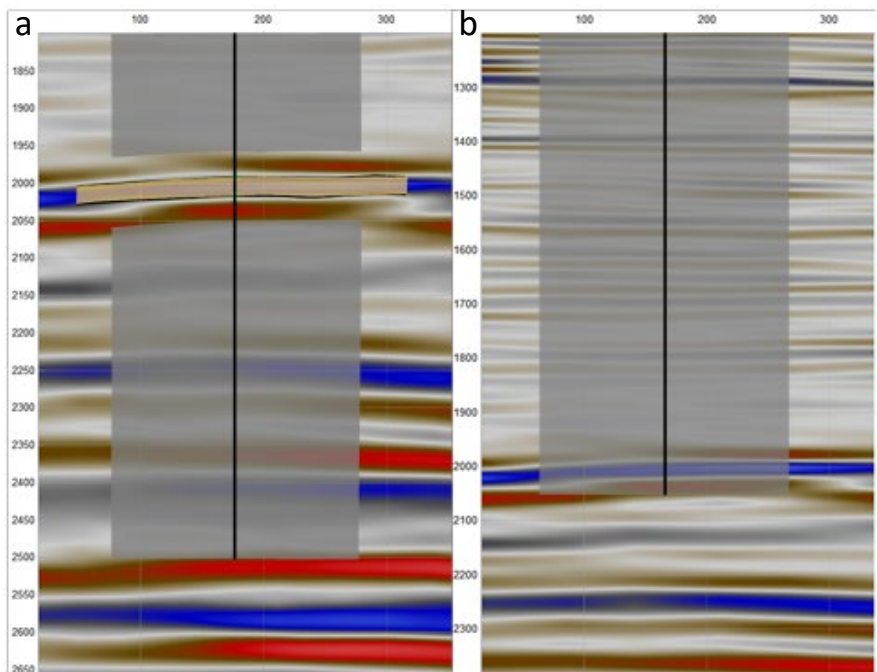

Figure 2: Example of an amplitude cube seismic section near wells with positive and negative class volumes allocated. The left section (a) is an example of class allocation for a productive well. Right section (b) is an example for a dry well. Yellow area - positive class, gray area - negative class.

**Creating features and dataset for training**

Using 3D seismic data for each seismic cube, 456 seismic wavefield attributes were obtained by applying 25 different attribute extraction algorithms with different window sizes: 10, 20, 30 and 50. The spectral decomposition algorithm was used to create mono-frequency cubes in the frequency channel range from 1 to 45 Hz with frequency decomposition by Ricker and Morlet wavelets. The attributes of the space were standardized and a Yeo-Johnson stepwise transformation was performed. Each point in space, within the extracted class volumes, was assigned a class and a vector of 1368 seismic attributes.

The selection of features for further training was carried out in two stages. At the first stage, the BoostARoota algorithm [3] was used. An algorithm from the CatBoost gradient-based decision tree (GBDT) family [4] was used as an estimator for its operation. In the second step, feature importance estimation was performed by training the same algorithm, but with noise added to the dataset and calculating the Shepley index for the features. The intentionally random data (noise) forms a threshold value on the Shepley index. All the features having index values less than the threshold value are excluded from the dataset. After two stages of selection, 87 features of the study space are left in the training dataset out of 1368.

The data were divided into training, validation and test samples. Two wells were known to have produced commercial gas flows, but did not have any logging data. Well SD-12 East was known to have encountered a separate gas accumulation in the basal KES, and well SD-5X penetrated productive sediments in the SD field. These wells were used in the study as test wells. Detailed correlation of the productive interval using seismic data was performed for these wells, and 50 meters was chosen as the radius of the gas zone based on minimum expert estimates. Also, to quantitatively assess the quality of the forecast, class labels of well IY-1X were excluded from the training process. The data on these wells were used as test wells for the obtained metamodel after the final prediction of the probability space of gas traps presence. Thus, a test sample was formed, which included 3 wells (Table 2).

Table 2. Class label sets for training and test

| Class label | Train+Validation | Test | Sum |
|---|---|---|---|
| Class 0 | 510501 | 16723 | 527 224 |
| Class 1 | 17595 | 2955 | 20 550 |

The total number of labels for the positive class (Class 1) was more than 20 thousand, and for the negative class (Class 0) - more than half a million, which allows to involve in training an order of magnitude more information about the space than on datasets where the training sample is formed using the classification of lithologic logging along the wellbore, for example, in [5] for prediction used a basic training sample of only 1127 labels. Increasing the size of the labeled data makes this approach more effective for training algorithms, as there is less chance of overtraining when using large amounts of data. Models are less likely to remember noise in the data and more likely to extract general patterns. Also, large amounts of data help improve the stability and reliability of models. The larger the data, the more complex dependencies between model inputs and outputs can be discovered. Models with larger data can better model nonlinear and complex relationships, which is especially useful in tasks that require a high degree of complexity. Bigger data helps the model to better interpret context and generalize knowledge to new situations.

The training and validation dataset consisted of drilling and testing results from 9 wells, of which 6 wells produced commercial gas flows from productive sediments and 3 wells produced no flows. Test wells were used for validation of machine learning algorithms. Thus, 8 datasets were generated for training and validation of basic prediction models. For each set, two test wells were sequentially removed from the training sample, on which the validation of the algorithms was performed, and training was performed on the remaining wells.

**Training of algorithms and evaluation of prediction quality**

The training process was conducted in two iterations. In the first iteration, the basic models were trained, and then the metamodel was trained based on their prediction. In this iteration, the quality of the prediction was evaluated using a test data set. In the second iteration, the test data were returned to the training sample, and the already trained models were further trained for a fixed number of epochs. Thus, the final model "saw" the entire available dataset for the study area.

The baseline models are derived from the results of three gradient-based bousting over decision trees (GBDT) algorithms CatBoost [4], LightGBM [6] and XGboost [7] on 8 datasets. For all machine learning algorithms, hyperparameter optimization was performed by maximizing the ROC AUC metric on the validation dataset with balancing the class ratio. The hyperparameters were selected using the TPESampler optimizer [8, 9]. Training was monitored by estimating the f1 score on the validation sample. The total number of trained models was 24. Further, these models were ensembled by the metamodel by stacking method based on logistic regression, and the quality of classification by the f1 score metric was evaluated by the test sample (Table 3).

Table 3: Values of the f1 score metric

| Class | Mean base models | Metamodel |
|---|---|---|
| 0 | 0.989489 | 0.998622 |
| 1 | 0.855921 | 0.893846 |

Table 3 shows the f1 score metric values for the metamodel and the arithmetic mean values for the base models. There is a significant improvement in the prediction of gas sandstone, represented by class 1, after ensemble of models by stacking method. The final values obtained for both classes indicate a high quality of prediction of the division of space into productive (gas saturated) and nonproductive using seismic attributes.

**Evaluating the contribution of seismic attributes to prediction**

Using a population of gradient bousting models over decision trees, the importance of features for prediction was evaluated (Table 4). The evaluation was performed using the SHAP library [10] by calculating the Shapley index. The obtained results were scaled (min-max) and then summarized for each feature. Table 4 shows the 20 traits out of 87 with maximum total contribution. The first column of the table shows the abbreviated name of the seismic cube, the second column shows the attribute name, and the third column shows the rank by total contribution as a fraction of the maximum value. The greatest contribution to the prediction was made by the spectral decomposition attributes with Morlet wavelet, where the digit in the name is the decomposition mono-frequency in Hz. The second most important for the prediction was sweetness.

Table 4. Twenty most important features for GBDT algorithms

| Seis. cub | Metod | Rank |
|---|---|---|
| far | spectral dec. morlet 7 | 1 |
| far | spectral dec. morlet 8 | 0.506899 |
| far | sweetness | 0.502528 |
| mid | sweetness | 0.498849 |
| far | spectral dec. morlet 3 | 0.45401 |
| mid | spectral dec. morlet 4 | 0.377284 |
| far | spectral dec. morlet 5 | 0.316045 |
| mid | instantaneous amplitude | 0.314788 |
| far | instantaneous amplitude | 0.301863 |
| mid | spectral dec. morlet 3 | 0.28018 |
| far | spectral dec. morlet 4 | 0.251218 |
| mid | acoustic impedance | 0.250115 |
| far | spectral dec. morlet 12 | 0.238255 |
| far | spectral dec. morlet 10 | 0.233682 |
| far | spectral dec. morlet 13 | 0.230201 |
| mid | spectral dec. morlet 10 | 0.213501 |
| mid | spectral dec. morlet 3 | 0.204309 |
| mid | tgradient magnitude | 0.203342 |
| mid | spectral dec. morlet 15 | 0.202428 |
| mid | spectral dec. morlet 11 | 0.197109 |

**Prediction**

Based on the results of pre-training of the base models and their ensemble with subsequent calibration of the metamodel probabilities on isotonic regression, a prediction was made. A three-dimensional cube of calibrated probabilities of belonging of the studied space to the class of gas sandstones (Class 1) was obtained.

A map of predicted effective gas saturated thicknesses at depths of 200 meters up and down from the surface of the base of the KES Formation was calculated from a three-dimensional probability cube. The interval was chosen because of significant uncertainty in the correlations of the main seismic horizons, including the horizon associated with the base of the KES Formation. The predicted effective thicknesses were calculated using the following formula:

$$h_{ij} = \sum_{k=1}^{K} 1_{(P_{ijk} \geq P_T)} s ,$$

where $s$ vertical resolution of the approximation grid, $P_{ijk}$ predicted calibrated probability of voxel belonging to the "gas collector" class, $P_T$ threshold value of probability of voxel belonging to the "gas collector" class, $1_{(\cdot)}$ indicator function. A vertical approximation grid resolution of 5 metres per voxel was used to construct the map, and the probability threshold $P_T$ was set to 0.5 (Figure 3).

Most of the predicted gas traps are lithological (Figure 3b) and do not coincide in plan with the identified non-anticlinal structures, which are shown as black polygons in Figure 3a.

On the map of predicted gas saturated thicknesses, the contours of open fields SD and IY are well delineated. At the same time, relatively large gas traps are predicted in the study area in areas not yet covered by drilling wells. These traps are shown in Figure 3a and are located within +- 200 meters of the base of the KES Formation. Significant volumes of space with a high probability of being gas sands below the mapped interval (Figure 3b), confined to Cretaceous sediments, are noted.

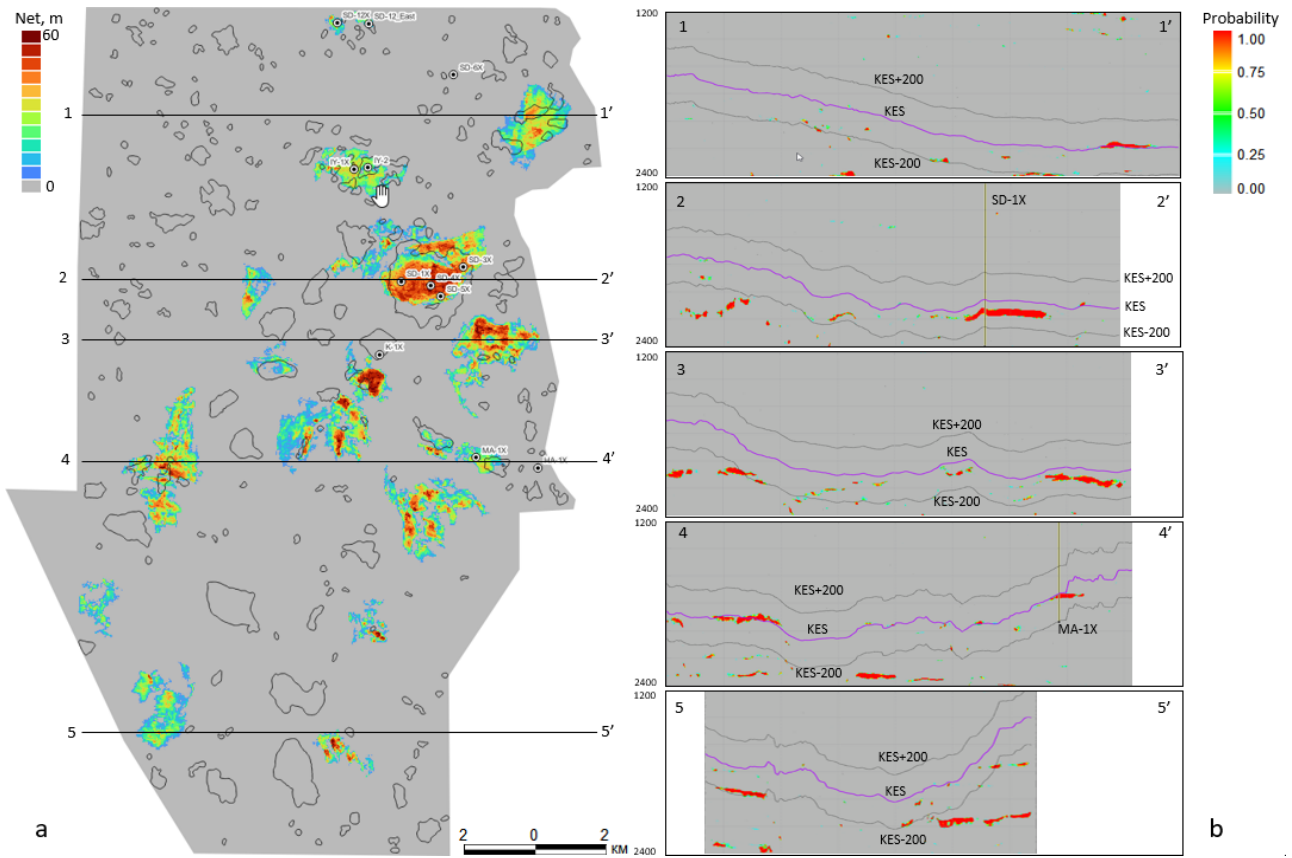

Figure 3: Left part (a) - predicted gas thicknesses in the depth interval +- 200 m from baseKES. Black polygons - anticlinal structures along the baseKES horizon along the deepest trailing isohypse. Right part(b) - sections through the cube of probabilities of space belonging to gas sands. Purple line - seismic horizon confined to the base of the KES Formation, black lines - horizons with 200 m offset up and down the section.

**Conclusions**

The paper formalizes an approach to creating a training dataset by selecting volumes with established gas saturation and filtration properties within a seismic wavefield. The volumes divide the studied space into positive and negative classes. Positive class is a volume of gas-saturated sands, identified by the results of detailed correlation of the gas sandstone top and bottom, within which there is a region with a boundary along the radius of continuous radial gas filtration. The radius was estimated by interpreting gas well test data after deconvolution of test regimes. The negative class includes volumes of explored space within a one-hundred-meter radius along the wellbore with proven futility based on test results and log interpretation. Each point in the space within the classes is assigned seismic field attributes. The training data set created in this way is used in the technological stack of sequential application of data processing methods and ensemble machine learning algorithms. As a result, the cube of calibrated probabilities of belonging of the studied space to gas reservoirs is obtained. The high efficiency of this approach is shown on a delayed test sample from three wells (blind wells). The final value of the gas sandstone prediction quality metric f1 score was 0.893846.

# Reference

1. Alain Gringarten From Straight Lines to Deconvolution: The Evolution of the State of the Art in Well Test Analysis 2008.
2. Van Poolen, H. Radius-of-drainage and stabilization-time equations. The Oil and Gas Journal, 1381964.
3. BoostARoota GitHub repository. https://github.com/chasedehan/BoostARoota.
4. Prokhorenkova G., Gusev A., Vorobev A., Dorogush, Gulin A. CatBoost: unbiased boosting with categorical features. In Bengio S., Wallach H., Larochelle H., Grauman K., Cesa-Bianchi N., Garnett R. editors. Proceedings of the 31st International Conference on Advances in Neural Information Processing Systems (NeurIPS'18). Curran Associates, 2018.
5. Ivlev D. Reservoir Prediction by Machine Learning Methods on The Well Data and Seismic Attributes for Complex Coastal Conditions arXiv preprint arXiv:2301.03216v1 2023.
6. Bergstra James, Daniel Yamins, David Cox. Making a science of model search: Hyperparameter optimization in hundreds of dimensions for vision architectures. Proceedings of The 30th International Conference on Machine Learning, 2013.
7. Ke Q. Meng, T. Finley, T. Wang, W. Chen, W. Ma, Q. Ye, T.-Y. Liu. Lightgbm: A highly efficient gradient boosting decision tree. In I. Guyon, U. von Luxburg, S. Bengio, H. Wallach, R. Fergus, S. Vishwanathan, and R. Garnett editors. Proceedings of the 30th International Conference on Advances in Neural Information Processing Systems (NeurIPS'17). Curran Associates, 2017.
8. Bergstra James S., et al. Algorithms for hyper-parameter optimization. Advances in Neural Information Processing Systems. 2011.
9. Chen and C. Guestrin. Xgboost: A scalable tree boosting system. In B. Krishnapuram, M. Shah, A. Smola, C. Aggarwal, D. Shen, and R. Rastogi editors. Proceedings of the 22nd ACM SIGKDD International Conference on Knowledge Discovery and Data Mining (KDD'16), ACM Press, 2016, pp. 785–794.
10. Scott Lundberg, Su-In Lee. A Unified Approach to Interpreting Model Predictions arXiv arXiv:1705.07874v2, 2017.